%% file: main.tex
\documentclass[journal]{IEEEtran}

\usepackage{graphicx}

\usepackage{amsmath,amsfonts,amssymb} 
\usepackage[ruled,vlined]{algorithm2e}
\usepackage{array}
\usepackage[caption=false,font=normalsize,labelfont=sf,textfont=sf]{subfig}
\usepackage{textcomp}
\usepackage{stfloats}
\usepackage{url}
\usepackage{verbatim}
\usepackage{cite}
\usepackage{xcolor}
\usepackage{balance}
\usepackage{glossaries}
\usepackage{ntheorem}   % for theorem-like environments
\usepackage{enumitem}

\def\BibTeX{{\rm B\kern-.05em{\sc i\kern-.025em b}\kern-.08em
    T\kern-.1667em\lower.7ex\hbox{E}\kern-.125emX}}
    
\DeclareMathOperator{\Tr}{Tr}
\newcommand\numberthis{\addtocounter{equation}{1}\tag{\theequation}} % To number equations
\DeclareMathOperator*{\argmax}{arg\,max}
\DeclareMathOperator*{\argmin}{arg\,min}

\begin{document}
\include{Heading/shortcuts}

\include{Heading/acronyms}
\newtheorem{assumption}{Assumption}
\newtheorem{lemma}{Lemma} 

\title{Orthogonal Least Squares with Integrated Information Theoretic Criteria for Joint Number of Targets and DoA Estimation}

\author{
Martin Willame,~\IEEEmembership{Graduated Student Member,~IEEE}, Gilles Monnoyer,~\IEEEmembership{Member,~IEEE},  \\
François Horlin,~\IEEEmembership{Fellow,~IEEE} and 
Jérôme Louveaux,~\IEEEmembership{Fellow,~IEEE}
        % <-this % stops a space
\thanks{This paper was submitted on the 07th of May, 2026. Martin Willame is with the Université Catholique de Louvain (UCLouvain) and the Université Libre de Bruxelles (ULB). Gilles Monnoyer and Jérôme Louveaux are with the UCLouvain (e-mails: $\{$martin.willame, gilles.monnoyer, jerome.louveaux@uclouvain.be$\}$). François Horlin is with the ULB (e-mails: francois.horlin@ulb.be$\}$).}% <-this % stops a space
}
% The paper headers
\markboth{Submitted to IEEE Signal Processing Letters}%
{Willame \MakeLowercase{\textit{et al.}}: Orthogonal Least Squares with Integrated Information Theoretic Criteria for Joint Number of Targets and DoA Estimation}

\maketitle

\begin{abstract}
We address the joint estimation of the number of targets and their \glspl{doa} using antenna arrays.
Target-number estimation can be formulated as a model-order selection problem and solved with the \gls{itc}. 
The \gls{itc} minimize an objective function that balances a likelihood term and a complexity penalty.
However, direct application of the \gls{itc} requires maximum-likelihood \gls{doa} estimates for each candidate model order, which is computationally prohibitive because it entails a multidimensional search over all angle combinations.
To reduce complexity, many radar processing exploit greedy methods such as \gls{ols}.
In this paper, we explore three distinct methods to integrate the \gls{itc} model-order selection into the \gls{ols} estimation procedure for joint target-number and \gls{doa} estimation. 
Specifically, we propose the disjoint rank-based, the joint selection-based, and the hybrid rank-and-selection-based \gls{itc}-\gls{ols} algorithms.
Each algorithm is derived under both the \gls{aic} and the \gls{bic} frameworks.
Numerical simulations show that the proposed hybrid \gls{itc}-\gls{ols} algorithm consistently outperforms both the other proposed variants and a baseline method from the literature.
\end{abstract}


% The text:
\input{Text/1-Introduction.tex}
\input{Text/2-System_Model.tex}
\input{Text/4-Joint.tex}

\input{Text/5-Simulations.tex}

\input{Text/6-Conclusion.tex}

\newpage
% The bibliography
\balance
\bibliographystyle{IEEEtran}
\bibliography{IEEEabrv,References}
\end{document}

%% file: Heading/shortcuts.tex
%% Comments
\newcommand{\GM}[1]{
{\color{blue} {\textbf{GM: }}
    #1
}}
\newcommand{\GMcomment}[1]{
{\footnotesize {\color{blue} {[\textbf{GM: }}
    #1
    ]
}}}

%% My maths symbols:

% Maths operators:
\newcommand{\defeq}{\mathrel{:=}}

% For Matrix:
\newcommand{\Diag}{\text{diag}}
\newcommand{\Vect}{\text{vec}}
\newcommand{\Transpose}[1]{{#1}^\mathrm{T}}
\newcommand{\Hermitian}[1]{{#1}^{\dagger}}
\newcommand{\MPInverse}[1]{{#1}^{+}}
\newcommand{\Inverse}[1]{{#1}^{-1}}
\newcommand{\Frob}{\mathrm{F}}
\newcommand{\Comp}{\bot}%{\mathsf{c}}
\newcommand{\SqrtMat}[1]{{#1}^{1/2}}
\newcommand{\ElemProd}{\odot}
\newcommand{\Kron}{\otimes}

% For variables:
\newcommand{\CandVal}[1]{\widetilde{#1}} % Candidate value
\newcommand{\EstVal}[1]{\widehat{#1}} % Estimated value
\newcommand{\RdmVal}[1]{\overline{#1}} % Random value

% For sets:
\newcommand{\Rset}{\mathbb{R}}
\newcommand{\Cset}{\mathbb{C}}

%% Parameters:

% Angle of arrival
\newcommand{\SRxInd}{m}
\newcommand{\SRxTh}{$m \textsuperscript{th}$ }
\newcommand{\SRxAll}{M}%\newcommand{\SRxAll}{N}
\newcommand{\Aoa}{\theta}
\newcommand{\AoaVec}{\boldsymbol \Theta}
\newcommand{\AoaSteerVec}{\mathbf{a}}
\newcommand{\AoaSteerMat}{\mathbf{A}}

% Subcarriers:
\newcommand{\SubInd}{q}
\newcommand{\SubTh}{$q \textsuperscript{th}$ }
\newcommand{\SubAll}{Q}
\newcommand{\Range}{\tau}
\newcommand{\RangeVec}{\bolsymbol \tau}
\newcommand{\RangeSteerVec}{\mathbf{t}}
\newcommand{\RangeSteerMat}{\mathbf{T}}

%% Targets:
\newcommand{\TgtInd}{k}
\newcommand{\TgtTh}{$k \textsuperscript{th}$ }
\newcommand{\TgtAll}{K}
\newcommand{\TgtPosVec}{\mathbf{x}}
\newcommand{\TgtPosMat}{\mathbf{X}}
\newcommand{\TgtPosX}{x}
\newcommand{\TgtPosY}{y}
\newcommand{\TgtVelVec}{\dot{\mathbf{x}}}
\newcommand{\TgtVelMat}{\dot{\mathbf{X}}}
\newcommand{\TgtVelX}{\dot{x}}
\newcommand{\TgtVelY}{\dot{y}}

% Observations:
\newcommand{\ObsInd}{d}
\newcommand{\ObsTh}{$d \textsuperscript{th}$ }
\newcommand{\ObsAll}{D}
\newcommand{\Doppler}{f}
\newcommand{\DopplerVec}{\boldsymbol \Omega}
\newcommand{\DopplerSteerVec}{\mathbf{f}}
\newcommand{\DopplerSteerMat}{\mathbf{F}}

% Radar Pairs:
\newcommand{\RpInd}{p}
\newcommand{\RpTh}{$p \textsuperscript{th}$ }
\newcommand{\RpAll}{P} 

%% Channel:
\newcommand{\ChnlVec}{\mathbf{h}}
\newcommand{\ChnlMat}{\mathbf{H}}
\newcommand{\MIMO}[1]{{#1}^\text{MIMO}}
\newcommand{\DiagVec}{\mathbf{d}}
\newcommand{\DiagMat}{\mathbf{D}}

% Channel coefficient:
\newcommand{\CoefVal}{\beta}
\newcommand{\CoefVec}{\boldsymbol \beta}
\newcommand{\CoefMat}{\mathbf{B}}
\newcommand{\CoefLength}{M}
\newcommand{\CoefVecbis}{\mathbf{b}}
\newcommand{\PathVal}{\alpha}
\newcommand{\PathVec}{\boldsymbol \alpha}

% noise value:
\newcommand{\NoiseVal}{n}
\newcommand{\NoiseVec}{\mathbf{n}}
\newcommand{\NoiseMat}{\mathbf{N}}
\newcommand{\NoiseVar}{\sigma^2}
\newcommand{\NoiseVarVec}{\boldsymbol \sigma^2}
\newcommand{\NoiseVarMM}{\sigma_c^2}
\newcommand{\NoiseSD}{\sigma}
\newcommand{\NoiseSDMM}{\sigma_c}
\newcommand{\NoiseSDMMp}{\sigma_{p,c}}
\newcommand{\SNR}{\mathrm{SNR}}

%% General matrices:
\newcommand{\SteerVec}{\mathbf{a}}
\newcommand{\SteerMat}{\mathbf{A}}
\newcommand{\SteerLength}{N}
\newcommand{\Identity}{\mathbf{I}}
\newcommand{\ProjMat}{\mathbf{P}}
\newcommand{\SCovMat}{\mathbf{R}}

%% Statistics:
\newcommand{\Pdf}{f}
\newcommand{\MeanVal}{\mu}
\newcommand{\MeanVec}{\boldsymbol \mu}
\newcommand{\CovMat}{\boldsymbol \Gamma}
\newcommand{\RdmVec}{\mathbf{z}}
\newcommand{\RdmMat}{\mathbf{Z}}
\newcommand{\CN}{\mathcal{CN}} % Complex Normal distribution

%% MLE:
\newcommand{\ParamVec}{\boldsymbol \gamma}
\newcommand{\LLFun}{\mathcal{L}}
\newcommand{\GridPoints}{n}
\newcommand{\Grid}{\mathcal{G}}
\newcommand{\FFTMat}{\mathbf{F}}
\newcommand{\Weigts}{w}

%% MUSIC:
\newcommand{\PseudoSpec}{\mathcal{J}}

\newcommand{\SigSpaceVec}{\mathbf{u}}
\newcommand{\SigSpaceMat}{\mathbf{U}}
\newcommand{\SigEigVal}{\lambda}
\newcommand{\SigEigVec}{\boldsymbol \lambda}
\newcommand{\SigEigMat}{\boldsymbol \Lambda}

\newcommand{\NoiseSpaceVec}{\mathbf{g}}
\newcommand{\NoiseSpaceMat}{\mathbf{G}}
\newcommand{\NoiseEigVal}{\varrho}
\newcommand{\NoiseEigVec}{\boldsymbol \varrho}
\newcommand{\NoiseEigMat}{\boldsymbol \Sigma}

%% Greedy algorithms:
\newcommand{\GreedySet}[2]{{#1}_{#2}}

%% OFDM:
\newcommand{\SubSpacing}{\Delta_f}
\newcommand{\Bandwith}{B}
\newcommand{\TimeFrame}{T}
\newcommand{\CpLen}{T_{\mathrm{cp}}}
\newcommand{\SymbolVal}{s}
\newcommand{\SymbolVec}{\mathbf{s}}
\newcommand{\SymbolMat}{\mathbf{S}}
\newcommand{\MappingMat}{\boldsymbol \Xi}
\newcommand{\SRxSigVec}{\mathbf{y}}
\newcommand{\SRxSigMat}{\mathbf{Y}}
\newcommand{\CarrierFreq}{f}
\newcommand{\Wavenumber}{k_c}
\newcommand{\SpeedOfLight}{c}

%% file: Heading/acronyms.tex
%% System Model:
\newacronym{ap}{AP}{access point}
\newacronym{pwr}{PWR}{passive Wi-Fi radar}
\newacronym{pr}{PR}{passive radar}
\newacronym{stx}{STx}{sensing transmitter}
\newacronym{srx}{SRx}{sensing receiver}
\newacronym{rp}{RP}{radar pair}
\newacronym{ula}{ULA}{uniform linear array}
\newacronym{simo}{SIMO}{single-input and multiple-output}
\newacronym{aoa}{AoA}{angle-of-arrival}
\newacronym{doa}{DoA}{direction-of-arrival}
\newacronym{snr}{SNR}{signal-to-noise ratio}
\newacronym{awgn}{AWGN}{additive white Gaussian noise}
\newacronym{ofdm}{OFDM}{orthogonal frequency-division multiplexing}
%% Algorithms
\newacronym{ols}{OLS}{orthogonal least squares}
\newacronym{omp}{OMP}{orthogonal matching pursuit}
\newacronym{ml}{ML}{maximum likelihood}
\newacronym{mle}{MLE}{maximum likelihood estimator}

%% ITCs:
\newacronym{itc}{ITC}{information theoretic criteria}
\newacronym{aic}{AIC}{Akaike information criterion}
\newacronym{aicc}{AICc}{corrected AIC}
\newacronym{bic}{BIC}{ Bayesian information criterion}
\newacronym{decitc}{Dec-ITC}{decision-based ITC}
\newacronym{eigitc}{Eig-ITC}{eigenvalue-based ITC}
\newacronym{cbnitc}{Cbn-ITC}{combined ITC}
\newacronym{decitcfull}{Dec-ITC}{decision-based information theoretic criteria}
\newacronym{eigitcfull}{Eig-ITC}{eigenvalue-based information theoretic criteria}
\newacronym{cbnitcfull}{Cbn-ITC}{combined information theoretic criteria}
\newacronym{decaic}{Dec-AIC}{decision-based AIC}
\newacronym{decbic}{Dec-BIC}{decision-based BIC}
\newacronym{eigaic}{Eig-AIC}{eigenvalue-based AIC}
\newacronym{eigbic}{Eig-BIC}{eigenvalue-based BIC}
\newacronym{cbnaic}{Cbn-AIC}{combined AIC}
\newacronym{cbnbic}{Cbn-BIC}{combined BIC}

%% Others
\newacronym{rmse}{RMSE}{root mean square error}
\newacronym{pdf}{PDF}{probability density function}
\newacronym{iid}{i.i.d.}{independent and identically distributed}
\newacronym{evd}{EVD}{eigen value decomposition}
\newacronym{svd}{SVD}{Singular Value Decomposition}
\newacronym{ff}{FF}{Far-Field}
\newacronym{crlb}{CRLB}{Cramér-Rao Lower Bound}

\newacronym{bols}{BOLS}{Block orthogonal least squares}
\newacronym{bomp}{BOMP}{Block orthogonal matching pursuit}
\newacronym{mimo}{MIMO}{Multiple-input and multiple-output}

%% file: Text/1-Introduction.tex
\section{Introduction} \label{sec:introduction}
\glsresetall
\IEEEPARstart{I}{n} many radar applications, antenna arrays are used to estimate the \glspl{doa} of multiple targets from noisy observations \cite{Richards2013PrinciplesOM,Foutz2008DoA,salama2025DoA,pesavento_three_2023}.
Although the \gls{ml} estimator can asymptotically attain the \gls{crlb}, its computational complexity grows exponentially with the number of targets; consequently, it is often unsuitable for real-time operation \cite{Kay1993FundamentalsOS}.
Low-complexity heuristics such as MUSIC are commonly used in \glspl{doa} estimation scenarios~\cite{1143830}. 
However, these subspace-based methods degrade when target signals are highly correlated \cite{pesavento_three_2023}.
Alternatively, greedy techniques like \gls{ols}, can be used as fast approximate solvers for the above \gls{ml} problem~\cite{chen_orthogonal_1989, blumensath_difference_2007, 10979421}.
This method iteratively build a solution by selecting one \gls{doa} at a time based on a least squares criterion.
This significantly reduces the computational burden compared to exhaustive search methods such as the \gls{ml} estimator.
In the litterature, the number of targets, which typically amounts to the number of steps in the greedy algorithm, is often assumed to be known \cite{pesavento_three_2023}.
However, in practice, it is usually an unknown parameter that must be estimated from the observed data.

Building upon this observation, this work addresses the joint estimation of the number of targets and their \glspl{doa} within an \gls{ols}-based framework.
A common strategy is to stop the greedy search using a residual-energy criterion \cite{5895106,LIANG2017165,12047,CHEN2019331}.
Yet fixed-threshold stopping remains difficult to tune in practice because the optimal threshold depends on the noise and received signal levels.
In contrast, within the \gls{ml} framework, target-number estimation is naturally formulated as a model-order selection problem.
It can be solved by the \gls{itc} that select the model minimizing an objective function that balances a likelihood term against a model complexity penalty \cite{Akaike1973InformationTA,Schwarz1978EstimatingTD,1164557,1311138,Neath2012BayesianIC,8498082,Cavanaugh2019AkaikeIT}.
The \gls{aic} and the \gls{bic} are two widely used \gls{itc} that differ by their model complexity penalty terms \cite{1311138}.
Nevertheless, a direct use of \gls{itc} requires \gls{ml} parameter estimation for every candidate model order, i.e., for each possible number of targets; this requirement again limits real-time applicability.

Given the identified gaps in the litterature, our key contributions
are the following:
\begin{itemize}
    \item We propose \textbf{three novel algorithms} that combine the \gls{itc} target-number estimation with the \gls{ols} greedy estimation of target \glspl{doa}: \textbf{disjoint, joint, and hybrid \gls{itc}-\gls{ols}}. Each algorithm is derived under both the \textbf{\gls{aic}} and the \textbf{\gls{bic}} formulations.
    \item We show that the need for \gls{ml} estimates in \gls{itc}-\gls{ols} can be circumvented by introducing an \textbf{\gls{ml}-correction parameter} in the \gls{itc} objective. This parameter \textbf{controls the tradeoff between hit rate and false-alarm rate at high \gls{snr}} while preserving performance in other regimes.
    \item The proposed \gls{itc}-\gls{ols} algorithms are applied in a \textbf{practical scenario} considering a \textbf{\gls{pr} setup} based on unknown \gls{ofdm} data frames. The \gls{pr} \textbf{estimates \glspl{doa} without decoding the transmitted symbols}.
    \item Through numerical simulations, we demonstrate that among the proposed algorithms, \textbf{hybrid \gls{itc}-\gls{ols}} provides the \textbf{best overall performance} across relevant radar parameters. We also show that, in this context, \textbf{the \gls{aic}-\gls{ols} consistently outperforms the \gls{bic}-\gls{ols}}.
\end{itemize}
The vectors and matrices are defined as $\mathbf{a}$ and $\mathbf{A}$, while $\mathbf{I}_\SRxAll$ denotes the identity matrix of size $\SRxAll$.
The real and complex sets are expressed as $\mathbb{R}$ and $\mathbb{C}$. 
The trace, the transpose and the Hermitian transpose are denoted $\Tr\left\{\mathbf{A}\right\}$ , $\Transpose{\mathbf{A}}$ and $\Hermitian{\mathbf{A}}$.

%% file: Text/2-System_Model.tex
\section{System Model} \label{sec:system_model}

We consider a \gls{pr} system designed to estimate the \gls{doa} of multiple targets using the reflected signals transmitted by an \gls{ap}.
The single-antenna \gls{ap} transmits an \gls{ofdm} downlink signal that the passive receiver captures using a \gls{ula} with $\SRxAll$ antennas.
We assume half-wavelength antenna spacing and orientation towards the coverage area. 
The \gls{ap} transmits a sequence of $\ObsAll$ \gls{ofdm} data symbols, separated by $\TimeFrame$ seconds.
Each \gls{ofdm} symbol contains $\SubAll$ subcarriers with uniform spacing $\SubSpacing$.
The first subcarrier has frequency $\CarrierFreq$ and corresponding wavenumber $\Wavenumber \triangleq 2\pi \CarrierFreq / \SpeedOfLight$, where $\SpeedOfLight$ denotes the speed of light. 
The transmitted complex data symbol for the \ObsTh \gls{ofdm} symbol ($\ObsInd \in \{0,\dots,\ObsAll-1\}$) and \SubTh subcarrier ($\SubInd \in \{0,\dots,\SubAll-1\}$) is denoted $\SymbolVal_{\ObsInd,\SubInd} \in \Cset$.
These symbols remain unknown to the passive receiver.
The \gls{pr} objective is to estimate the \gls{doa} of $\TgtAll$ targets within the coverage area.
We define the true \gls{doa} of the \TgtTh target as $\Aoa_\TgtInd \in [-\pi/2,\pi/2]$. 
\glspl{doa} measure the angle between the incoming wavefront and the normal vector of the receiver antenna array. 
Collecting all target \glspl{doa}, we define the \gls{doa} vector $\AoaVec=\Transpose{[\Aoa_1 \ \dots \ \Aoa_\TgtAll]} \in \Rset^{\TgtAll \times 1}$.
Under the far-field assumption for all targets with respect to the receiver, the \gls{doa} steering vector $\SteerVec \in \Cset^{\SRxAll \times 1}$ and the corresponding steering matrix $\SteerMat(\AoaVec) \in \Cset^{\SRxAll \times \TgtAll}$ are defined as
\begin{align}
    \SteerVec(\Aoa_\TgtInd) & = \Transpose{\left[1 \ e^{j\pi \sin(\Aoa_\TgtInd)} \dots \ e^{j\pi (\SRxAll-1) \sin(\Aoa_\TgtInd)}\right]}, \\
    \SteerMat(\AoaVec) & = [\SteerVec(\Aoa_1) \ \dots \ \SteerVec(\Aoa_\TgtAll)].
\end{align}
We ground the channel model on the following assumptions: first only single-bounce multipath signals contribute significantly to the observed channel; second the receiver achieves perfect timing and frequency synchronization using the transmitted preamble.
Under these assumptions, the channel for the \ObsTh \gls{ofdm} symbol and \SubTh subcarrier is expressed as $\ChnlVec_{\ObsInd,\SubInd}(\AoaVec,\CoefVec_{\ObsInd,\SubInd}) = \SteerMat(\AoaVec) \ \CoefVec_{\ObsInd,\SubInd}$, with, 
\begin{align}
    \CoefVec_{\ObsInd,\SubInd} & = \Transpose{[\CoefVal_{\ObsInd,\SubInd,1} \ \dots \ \CoefVal_{\ObsInd,\SubInd,\TgtAll}]}, \in \Cset^{\TgtAll \times 1}, \\
    \CoefVal_{\ObsInd,\SubInd,\TgtInd} & = \PathVal_{\TgtInd} \ e^{-j2\pi \Wavenumber \Range_{\TgtInd}} \ e^{-j2\pi \SubSpacing \Range_{\TgtInd} \SubInd} \ e^{j2\pi \Doppler_{\TgtInd} \ObsInd/\ObsAll \TimeFrame}, \in \Cset.
\end{align}
where the complex channel coefficient vector $\CoefVec_{\ObsInd,\SubInd}$ captures, for each target $\TgtInd$, the path loss, the radar cross-section and the propagation phase (all encoded in $\alpha_{\TgtInd} \in \Cset$), the range-induced phase shift ($\Range_{\TgtInd}$), and the Doppler shift ($\Doppler_{\TgtInd}$) \cite{9477585}.  
The model characterizing the received signal $\SRxSigVec_{\ObsInd,\SubInd} \in \Cset^{\SRxAll \times 1}$ at the \gls{pr} is given by 
\begin{equation} \label{eq:SRx_sig}
\SRxSigVec_{\ObsInd,\SubInd} = \ChnlVec_{\ObsInd,\SubInd}(\AoaVec,\CoefVec_{\ObsInd,\SubInd}) \ \SymbolVal_{\ObsInd,\SubInd} + \NoiseVec_{\ObsInd,\SubInd} = \SteerMat(\AoaVec) \ \CoefVec_{\ObsInd,\SubInd}^{\prime} + \NoiseVec_{\ObsInd,\SubInd},
\end{equation}
where $\CoefVec_{\ObsInd,\SubInd}^{\prime}\triangleq \CoefVec_{\ObsInd,\SubInd} \ \SymbolVal_{\ObsInd,\SubInd} \in \Cset^{\TgtAll \times 1}$ represents the channel coefficients modulated by the unknown data symbol, and $\NoiseVec_{\ObsInd,\SubInd} \in \Cset^{\SRxAll \times 1}$ denotes the \gls{awgn} contribution.
Each noise vector $\NoiseVec_{\ObsInd,\SubInd} \overset{\text{\acrshort{iid}}}{\sim} \CN(\mathbf{0}, \NoiseVar \ \Identity_\SRxAll)$ is an \gls{iid} circularly symmetric complex Gaussian random vector with zero mean and diagonal covariance matrix. 
The corresponding \gls{snr} is defined as
\begin{equation} \label{eq:snr}
    \SNR = \frac{\frac{1}{\TgtAll}\sum_{\TgtInd=1}^{\TgtAll} |\PathVal_\TgtInd |^2}{\NoiseVar}. 
\end{equation}

%% file: Text/4-Joint.tex
\section{Joint Target Number and DoA Estimation} \label{sec:joint}
This section presents three algorithms that combine the \gls{itc} for target-number detection with the \gls{ols} greedy estimation of target \glspl{doa}.
In this context, \gls{itc} estimate $\TgtAll$ by minimizing an objective function balancing two terms: one likelihood term that quantifies the fit of the model to the observed data, and one model complexity penalty term that discourages overfitting by penalizing a higher number of targets.
In this work, the term \gls{itc} refers to both the \gls{aic} and the \gls{bic} criteria, which differ by their model complexity penalty terms. 
The three approaches are detailed in the following subsections, and their performance is compared in Section~\ref{sec:sim_res}. 

\subsection{Disjoint rank-based \gls{itc}-\gls{ols}}
In this disjoint approach, the number of targets is first estimated using rank-based \gls{itc}, agnostically of the \gls{doa} estimator.
Building upon this estimate, \gls{ols} is then used to estimate the target \glspl{doa} for the inferred target number.

\subsubsection{\textbf{Target Number Estimation with Rank-based \gls{itc}}} \label{sec:rank_itc}
The rank-based \gls{itc} estimates the number of targets $\TgtAll$ by analyzing the rank of the sample covariance matrix \cite{1164557}, 
% when $\ObsAll \SubAll > \SRxAll$ \cite{1164557},
\begin{equation}
\SCovMat \textstyle=\frac{1}{\ObsAll\SubAll} \sum_{\ObsInd=0}^{\ObsAll-1} \sum_{\SubInd=0}^{\SubAll-1} \SRxSigVec_{\ObsInd,\SubInd} \ \Hermitian{\SRxSigVec}_{\ObsInd,\SubInd}, \in \Cset^{\SRxAll \times \SRxAll}.
\end{equation}
Letting $\lambda_1 \geq \lambda_2 \geq \dots \geq \lambda_\SRxAll$ be the ordered eigenvalues of $\SCovMat$, the rank is estimated as
$\EstVal{\TgtAll} = \argmin_{\CandVal{\TgtInd}} \text{ITC}^{\text{Rank}} (\CandVal{\TgtInd})$.
The discrete function of the rank $\text{ITC}^{\text{Rank}}$ is defined as
\begin{equation} \label{eq:itc_rank}
    \text{ITC}^{\text{Rank}} (\CandVal{\TgtInd}) = -C(\CandVal{\TgtInd}) \log\left(\frac{\prod_{i=\CandVal{\TgtInd}+1}^{\SRxAll} \lambda_i^{1/(\SRxAll-\CandVal{\TgtInd})}}{\frac{1}{(\SRxAll-\CandVal{\TgtInd})}\sum_{i=\CandVal{\TgtInd}+1}^{\SRxAll} \lambda_i} \right) + \ \nu(\CandVal{\TgtInd}),
\end{equation}
with $C(\CandVal{\TgtInd}) = 2\ObsAll\SubAll(\SRxAll-\CandVal{\TgtInd})$, and with the model complexity penalty term $\nu(\CandVal{\TgtInd})$ defined as $\nu_{\text{AIC}}(\CandVal{\TgtInd}) = 2\CandVal{\TgtInd}(2\SRxAll - \CandVal{\TgtInd})$ for \gls{aic} and $\nu_{\text{BIC}}(\CandVal{\TgtInd}) = \CandVal{\TgtInd}(2\SRxAll - \CandVal{\TgtInd})\log(\ObsAll\SubAll)$ for \gls{bic}. 
The likelihood term in \eqref{eq:itc_rank} evaluates how well the observed sample covariance matrix $\SCovMat$ fits with a rank $\CandVal{\TgtInd}$ matrix, while $\nu(\CandVal{\TgtInd})$ penalizes higher $\CandVal{\TgtInd}$, corresponding to a higher number of targets.

\subsubsection{\textbf{\gls{doa} Estimation with \gls{ols}}} \label{sec:OLS}
\begin{algorithm}[t]
    \caption{\gls{ols} with $\EstVal{\TgtAll}$ Iterations}
    \label{alg:ols}
    \KwIn{$\EstVal{\TgtAll}$ and $\SRxSigVec_{\ObsInd,\SubInd},  \forall \ObsInd, \SubInd$}
    \vspace{0.05cm}
    \KwOut{$\widehat{\AoaVec}_{\EstVal{\TgtAll}} = [\widehat{\Aoa}_1 \ \dots  \ \widehat{\Aoa}_{\EstVal{\TgtAll}}]$}

    \Begin{
    \textbf{i. Initialization:} $\GreedySet{\EstVal{\AoaVec}}{0} \gets [~]$, $\ProjMat^{\Comp}\big(\GreedySet{\EstVal{\AoaVec}}{0}\big) \gets \Identity_\SRxAll$\;
        \For{$k\gets 0$ \KwTo $\EstVal{\TgtAll}-1$}{
        	\textbf{ii. Selection:} $~\EstVal{\Aoa}_{\TgtInd+1} \gets $ \eqref{eq:simo_OLS_R}\;
        	\textbf{iii. Update:} $\GreedySet{\EstVal{\AoaVec}}{\TgtInd+1} \gets [\GreedySet{\EstVal{\AoaVec}}{\TgtInd},\EstVal{\Aoa}_{\TgtInd+1}],$ $\ProjMat^{\Comp}\big(\GreedySet{\EstVal{\AoaVec}}{\TgtInd+1}\big)\gets $ \eqref{eq:simo_CompProjMat}\;
    	}
    }
\end{algorithm}
Algorithm~\ref{alg:ols} outlines the successive steps of \gls{ols}, which we detail next.
We denote the estimated angle set at iteration $k$ as $\GreedySet{\EstVal{\AoaVec}}{k} = [\EstVal{\Aoa}_1 \ \dots \ \EstVal{\Aoa}_k]$.
The projection matrices spanning the subspace generated by the steering matrix $\SteerMat(\GreedySet{\EstVal{\AoaVec}}{k})$ and its orthogonal complement are defined as 
\begin{align}
    & \ProjMat\big(\GreedySet{\EstVal{\AoaVec}}{\TgtInd}\big) = \AoaSteerMat\big(\GreedySet{\EstVal{\AoaVec}}{\TgtInd}\big) \Inverse{\big(\Hermitian{\AoaSteerMat}\big(\GreedySet{\EstVal{\AoaVec}}{\TgtInd}\big) \AoaSteerMat\big(\GreedySet{\EstVal{\AoaVec}}{\TgtInd}\big)\big)} \Hermitian{\AoaSteerMat}\big(\GreedySet{\EstVal{\AoaVec}}{\TgtInd}\big), \\
    & \ProjMat^{\Comp}\big(\GreedySet{\EstVal{\AoaVec}}{\TgtInd}\big) = \Identity_\SRxAll - \ProjMat\big(\GreedySet{\EstVal{\AoaVec}}{\TgtInd}\big). \label{eq:simo_CompProjMat}
\end{align}

\begin{enumerate} [label=\roman*.,wide, labelindent=0pt]
\item \textbf{Initialization:} This algorithm initializes with an empty angle set $\GreedySet{\EstVal{\AoaVec}}{0}$.
At initialization, $\ProjMat^{\Comp}\big(\GreedySet{\EstVal{\AoaVec}}{0}\big) = \Identity_\SRxAll$, which spans the full observation space.
The algorithm then alternates between selection and update steps until $\EstVal{\TgtAll}$ targets are estimated.
\item \textbf{Selection:} Each iteration selects the next target \gls{doa} $\EstVal{\Aoa}_{\TgtInd+1}$ by solving an optimization problem over a search grid,
\begin{align*}
    \EstVal{\Aoa}_{\TgtInd+1} 
        & = \argmin_{\CandVal{\Aoa}_{\TgtInd+1}} \sum_{\ObsInd=0}^{\ObsAll-1} \sum_{\SubInd=0}^{\SubAll-1} \min_{\CandVal{\CoefVec}_{\ObsInd,\SubInd}^{\prime}} \big\lVert \SRxSigVec_{\ObsInd,\SubInd} - \SteerMat(\GreedySet{\CandVal{\AoaVec}}{k+1}) \ \CandVal{\CoefVec}_{\ObsInd,\SubInd}^{\prime}  \big\rVert_2^2, \\
        & = \argmin_{\CandVal{\Aoa}_{\TgtInd+1}} \Tr\left[ \ProjMat^{\Comp}\big(\GreedySet{\CandVal{\AoaVec}}{k+1}\big) \ \SCovMat \right],  \\
        & = \argmax_{\CandVal{\Aoa}_{\TgtInd+1}} \ 
        \frac{\Hermitian{\SteerVec}(\CandVal{\Aoa}_{\TgtInd+1}) \ \GreedySet{\SCovMat}{\TgtInd} \ \SteerVec(\CandVal{\Aoa}_{\TgtInd+1})}{\lVert \Hermitian{\SteerVec}(\CandVal{\Aoa}_{\TgtInd+1}) \ProjMat^{\Comp} \big(\GreedySet{\EstVal{\AoaVec}}{\TgtInd}\big) \rVert_2^2}, \numberthis \label{eq:simo_OLS_R}
\end{align*}
in which the joint set composed of the previously estimated \glspl{doa} $\GreedySet{\EstVal{\AoaVec}}{k}$ and of the candidate \gls{doa} $\CandVal{\Aoa}_{k+1}$ is denoted by $\GreedySet{\CandVal{\AoaVec}}{k+1} = [\GreedySet{\EstVal{\AoaVec}}{k}, \CandVal{\Aoa}_{k+1}]$. The residual sample covariance at step $\TgtInd$ is given by $\GreedySet{\SCovMat}{\TgtInd} = \ProjMat^{\Comp}\big(\GreedySet{\EstVal{\AoaVec}}{\TgtInd}\big)\,\SCovMat\,\ProjMat^{\Comp}\big(\GreedySet{\EstVal{\AoaVec}}{\TgtInd}\big)$.
Equation \eqref{eq:simo_OLS_R} follows from a projection-matrix decomposition \cite{7543,10979421}.
As a result, this formulation avoids computing the full projection matrix $\ProjMat\big(\GreedySet{\CandVal{\AoaVec}}{k+1}\big)$ for each candidate angle, which significantly reduces the computational complexity of the selection step.
\item \textbf{Update:} The set of estimated angles $\GreedySet{\EstVal{\AoaVec}}{\TgtInd+1}$ and its corresponding orthogonal complement projection matrix in \eqref{eq:simo_CompProjMat} are updated to include the newly estimated angle $\EstVal{\Aoa}_{\TgtInd+1}$.
\end{enumerate}

\subsection{Joint Selection-based \gls{itc}-\gls{ols}} \label{sec:selection_itc_ols}
In contrast to the decoupled rank-based \gls{itc} approach, prior works have derived selection-based \gls{itc} that depend on the \gls{ml} estimates of the target \glspl{doa} $\GreedySet{\EstVal{\AoaVec}}{\TgtInd}^\text{ML}$ for each candidate model order $\TgtInd$ \cite{1311138,salama2025DoA}.
These \gls{itc} criteria have a likelihood term that evaluates the fit of the model defined in \eqref{eq:SRx_sig} for $\TgtInd$ targets to the received signal vectors $\SRxSigVec_{\ObsInd,\SubInd}$, penalized by a model complexity term as follows:
\begin{equation} \label{eq:itc_ml} 
    \text{ITC}^{\text{\gls{ml}}}(k) = \frac{-2}{\NoiseVar} \Tr\left[ \ProjMat^{\Comp}\big( \GreedySet{\EstVal{\AoaVec}}{\TgtInd}^\text{ML} \big) \ \SCovMat \right] + \ \eta(\TgtInd).
\end{equation}
The penalties are defined as $\eta_{\text{AIC}}(\TgtInd) = 2\TgtInd(1 + 2\ObsAll\SubAll)$ for \gls{aic} and $\eta_{\text{BIC}}(\TgtInd) = \TgtInd(1 + 2\ObsAll\SubAll)\log(\ObsAll\SubAll)$ for \gls{bic}.
However, computing \gls{ml} estimates for every candidate number of targets is computationally prohibitive.

To overcome this issue, we propose to use the greedy \gls{ols} algorithm to provide \gls{doa} estimate and introduce a modified \gls{itc} criterion within the \gls{ols} iterations.
This is achieved by conditioning the update step in Algorithm~\ref{alg:ols} on the value of the following proposed \gls{itc}-\gls{ols} objective at each iteration $\TgtInd$,  
\begin{equation} \label{eq:itc_ols}
    \text{ITC}^{\text{\gls{ols}}}(\TgtInd) = \frac{-2}{\max(\NoiseVar, \NoiseVarMM)} \Tr\left[ \ProjMat^{\Comp}\big(\GreedySet{\EstVal{\AoaVec}}{\TgtInd} \big) \ \SCovMat \right] + \ \eta(\TgtInd).
\end{equation}
Because \gls{ols} estimates $\GreedySet{\EstVal{\AoaVec}}{\TgtInd}$ are available at each step rather than \gls{ml} estimates $\GreedySet{\EstVal{\AoaVec}}{\TgtInd}^\text{ML}$, the criteria in \eqref{eq:itc_ols} include an \gls{ml} correction parameter $\NoiseVarMM$, which is interpreted as a minimum noise-variance level.
In the \gls{ols} workflow, instead of systematically adding the next estimated angle $\EstVal{\Aoa}_{\TgtInd+1}$ to $\GreedySet{\EstVal{\AoaVec}}{\TgtInd}$, we add $\EstVal{\Aoa}_{\TgtInd+1}$ only if it improves the objective, i.e., 
\begin{equation} \label{eq:stop_crit}
    \text{ITC}^{\text{\gls{ols}}}(\TgtInd+1) < \text{ITC}^{\text{\gls{ols}}}(\TgtInd).
\end{equation}
This acts as a stopping criterion for the \gls{itc}-\gls{ols} iterations.
At low \glspl{snr}, $\NoiseVarMM$ in \eqref{eq:itc_ols} has no impact because $\NoiseVar$ dominates the denominator.
At very high \glspl{snr}, i.e., as $\NoiseVar \to 0$, \gls{ml} estimates almost perfectly fit the ground truth, and the residual term $\Tr\left[ \ProjMat^{\Comp}\big( \GreedySet{\EstVal{\AoaVec}}{\TgtInd}^\text{ML} \big)\SCovMat \right]$ tends to zero for $\TgtInd = \TgtAll$.
For greedy \gls{ols} estimates, however, this residual term does not necessarily vanish as $\NoiseVar \to 0$.
Consequently, as this term is strongly amplified by the denominator in \eqref{eq:itc_ols}, the stopping criterion in \eqref{eq:stop_crit} can only be triggered for large penalty term (i.e. large $\TgtInd$). 
This leads to overfitting by introducing spurious targets close to already detected ones. 
Therefore, at high \glspl{snr}, $\NoiseVarMM$ provides a trade-off between detection rate and false-alarm rate.
Suggested values of $\NoiseVarMM$ are discussed in Section~\ref{sec:sim_res}.

\subsection{Hybrid Rank and Selection-based \gls{itc}-\gls{ols}} \label{sec:combined_itc_ols}
The selection-based \gls{itc}-\gls{ols} method can be further improved by combining it with rank-based \gls{itc}.
First, before running \gls{ols},the rank of the sample covariance matrix $\SCovMat$ is estimated to provide a preliminary estimate $\EstVal{\TgtAll}$ using the criterion in \eqref{eq:itc_rank}.
Next, during the first $\EstVal{\TgtAll}$ iterations of \gls{ols}, the estimated angles $\EstVal{\Aoa}_{\TgtInd+1}$ are systematically added to $\GreedySet{\EstVal{\AoaVec}}{\TgtInd}$.
As a result, the algorithm enforces a minimum number of selected targets determined by the estimated rank of $\SCovMat$.
For subsequent steps, i.e., when $\TgtInd+1 > \EstVal{\TgtAll}$, the newly estimated angle $\EstVal{\Aoa}_{\TgtInd+1}$ is added only if it improves the selection-based \gls{itc}-\gls{ols} objective in \eqref{eq:itc_ols}; therefore enabling the detection of more than $\EstVal{\TgtAll}$ targets. 
This combination is motivated by complementary operating regimes. 
At low \glspl{snr}, selection-based \gls{itc}-\gls{ols} tend to underestimate the number of targets due to the division by $\NoiseVar$ in \eqref{eq:itc_ols}. 
Rank-based \gls{itc} are typically more robust to noise and provide a reliable, yet conservative, target-count estimate.
At medium \glspl{snr}, selection-based \gls{itc}-\gls{ols} can improve the rank-based conservative target-count estimate by leveraging accurate \gls{ols} angle estimates. 
At high \glspl{snr}, the \gls{ml} correction parameter $\NoiseVarMM$ in \eqref{eq:itc_ols} can be tuned to control the tradeoff between hit and false-alarm rates. 
Therefore preventing overestimation of the number of targets.

%% file: Text/5-Simulations.tex
\section{Simulation Results} \label{sec:sim_res}
\begin{figure*}[!t]
\captionsetup[subfigure]{labelformat=empty, labelsep=none, justification=raggedright, singlelinecheck=false}

\hspace{0.05\textwidth}
\begin{center}
\subfloat[]{
\includegraphics[trim=3.4cm 9.5cm 4.3cm 9.91cm,width = 0.3\textwidth]{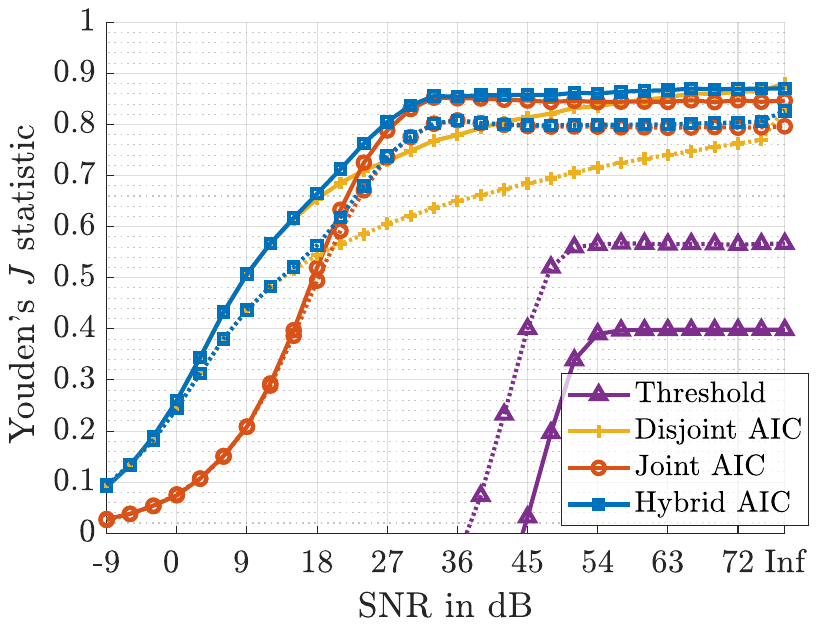}\label{fig:Jstat_vs_SNR}
}
\hfill
\subfloat[]{
\includegraphics[trim=3.4cm 9.5cm 4.3cm 9.91cm,width = 0.3\textwidth]{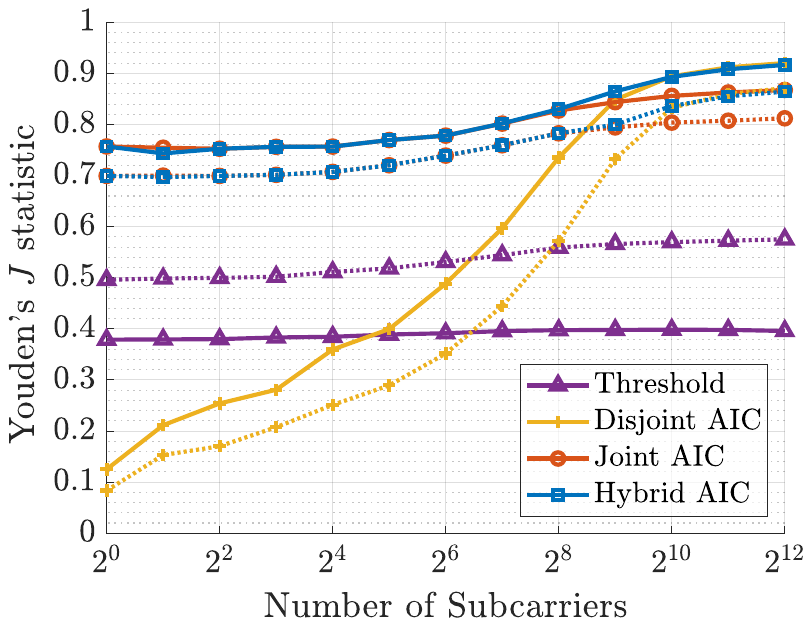}\label{fig:Jstat_vs_Q}
}
\hfill
\subfloat[]{
\includegraphics[trim=3.4cm 9.5cm 4.3cm 9.91cm,width = 0.3\textwidth]{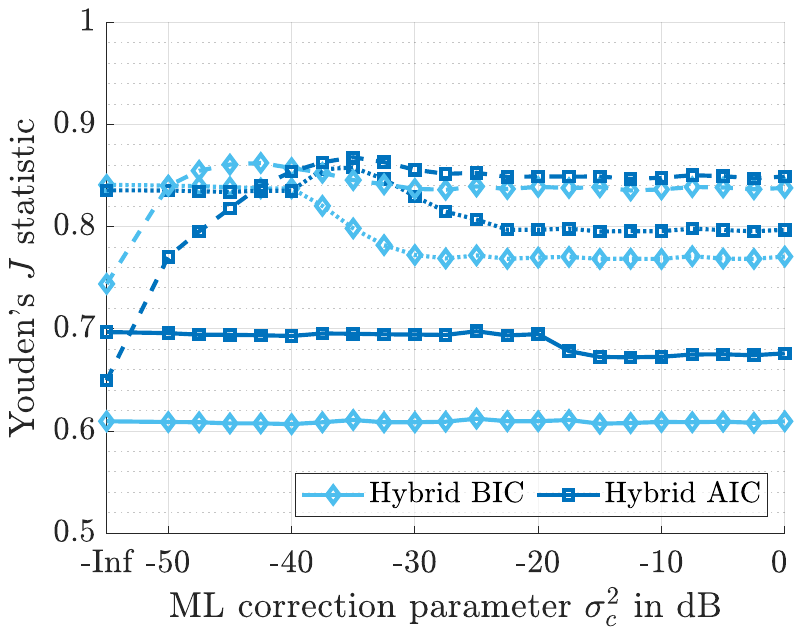}\label{fig:Jstat_vs_BNV}
}
\end{center}
\vspace*{-0.27\textwidth}
\hspace{-.01\textwidth} (a) \hspace{.305\textwidth} (b) \hspace{.305\textwidth} (c) \hfill
\vspace*{0.21\textwidth}
\caption{Comparison of the performance of the different \gls{itc}-\gls{ols} algorithms in terms of the Youden's $J$ statistics. (a) and (b) display the \gls{aic}-\gls{ols} as a function of the \gls{snr} and the number of targets $Q$, respectively, with solid and dotted lines corresponding to scenarios with $K=8$ and $K=12$, respectively. (c) compares the hybrid \gls{aic}-\gls{ols} and \gls{bic}-\gls{ols} algorithms as a function of the \gls{ml} correction parameter $\NoiseVarMM$, with solid, dotted and dashed lines corresponding to \glspl{snr} of $20$ dB, $40$ dB and $60$ dB, respectively.}
\label{fig:Jstat_comparison}
\end{figure*}

% Intro:
This section compares the proposed disjoint, joint, and hybrid \gls{itc}-\gls{ols} algorithms for jointly estimating the number of targets and their corresponding \glspl{doa}.
As a baseline, we consider the fixed-threshold stopping rule applied to the residual term $\Tr\left[ \ProjMat^{\Comp}\big( \GreedySet{\EstVal{\AoaVec}}{\TgtInd} \big) \ \SCovMat \right]$, which is commonly used in the literature \cite{CHEN2019331}.
This threshold is defined as $\NoiseVar (N_{obs}\sqrt{2 N_{obs} \ln{N_{obs}}})$ where $N_{obs} = \ObsAll \SubAll \SRxAll$ is the total number of observations.
Performance is evaluated using Youden's $J$ statistic, defined as the difference between hit rate and false-alarm rate \cite{Youden1950Index,MartinezCamblor2019}.
A score of $1$ indicates perfect detection, whereas a score of $0$ indicates equal hit and false-alarm rates.
A hit is declared when a detected peak falls within the main lobe of its associated true target \gls{doa}.
Conversely, a false alarm is declared when a detected peak cannot be associated with any true target \gls{doa}.
All metrics are computed over $10,000$ Monte-Carlo runs, in which detected \glspl{doa} are associated with true \glspl{doa} using the Hungarian algorithm \cite{Kuhn2010}.

For each simulation, we consider a fixed setup with colocated \gls{ap} and \gls{pr}, and with multiple targets randomly positioned within a maximum range of $\Range_{max} = 60$ m.
In accordance with the Wi-Fi 7 standard, the carrier frequency is set to $f_c = 5$ GHz and the subcarrier spacing to $\SubSpacing = 78.125$ kHz \cite{11090080}.
Unless otherwise stated, the parameters are fixed to $\TgtInd = 8$ targets, $\SRxAll = 16$ antennas, $Q = 512$ subcarriers (i.e., $40$ MHz), $\ObsAll = 10$ symbols, and $\SNR = 60$ dB.
The mean \gls{snr} is computed from \eqref{eq:snr} across all Monte-Carlo runs.
Because this definition is dominated by the strongest target, which is the easiest to detect, the effective radar operating \gls{snr} may appear higher than that of a conventional communication system.
\balance

The results are shown in \figurename~\ref{fig:Jstat_comparison}.
Unless otherwise stated, we use the \gls{aic} penalty and set the \gls{ml} correction parameter to $\NoiseVarMM = -35$ dB.
Overall, all proposed \gls{itc}-\gls{ols} variants outperform the fixed-threshold method. 
% Jstat vs SNR:
\figurename~\ref{fig:Jstat_vs_SNR} compares the performance as a function of the \gls{snr}. 
As expected, at low \gls{snr}, the joint selection-based \gls{aic}-\gls{ols} triggers the stopping criterion in \eqref{eq:stop_crit} too early because of the large division by $\NoiseVar$ in \eqref{eq:itc_ols}, leading to a low hit rate.
Consequently, in this regime, the disjoint rank-based \gls{aic} outperforms the joint approach.
At medium \gls{snr}, the trend reverses: as the \gls{snr} increases, detected peaks are more likely to correspond to true targets, improving the hit rate of the joint \gls{aic}-\gls{ols}, whereas the disjoint method remains agnostic to estimates quality.
At high \gls{snr}, both methods are reliable when the \gls{ml} correction parameter is properly tuned.
Across the full \gls{snr} range, the hybrid \gls{aic}-\gls{ols} method achieves the best performance by leveraging the complementarity of the two strategies. 

% Jstat vs Q:
\figurename~\ref{fig:Jstat_vs_Q} depicts the performance as a function of the number of subcarriers $Q$, while maintaining a constant sub-spacing $\SubSpacing$.
This is equivalent to illustrating the performance as a function of the bandwidth.
The performance of the disjoint rank-based \gls{aic}-\gls{ols} improves significantly as $Q$ increases because target echos become less correlated.
With the increasing bandwidth, the rank-based \gls{aic} discriminate more effectively between signal and noise subspaces.
The joint \gls{aic}-\gls{ols} method is less sensitive to $Q$ because it relies primarily on detected peaks rather than on explicit subspace separation.
Nevertheless, increasing $Q$ still improves its performance by providing more observations.
Once again, the hybrid \gls{aic}-\gls{ols} method outperforms both disjoint and joint methods across the full $Q$ range by combining their strengths. 

% Jstat vs BNV:
\figurename~\ref{fig:Jstat_vs_BNV} compares the performance of the hybrid \gls{aic} and \gls{bic} methods, as a function of the \gls{ml} correction parameter $\NoiseVarMM$, for different \glspl{snr}.
At low \gls{snr}, hybrid-method performance is mostly insensitive to $\NoiseVarMM$ because the rank-based \gls{itc} dominate.
In this regime, \gls{aic} outperforms \gls{bic} because its lower penalty term typically yields a higher hit rate.
At medium and high \gls{snr}, however, $\NoiseVarMM$ has a pronounced impact and must be tuned to balance hit and false-alarm rates.
As expected, when no correction parameter is used (i.e., $\NoiseVarMM = -\infty$ dB), the hybrid \gls{itc}-\gls{ols} performance degrades at high \gls{snr} because the selection-based \gls{itc} tend to generate more false alarms.
Across the tested parameters range, the operating values of $\NoiseVarMM$ that provide a favorable hit/false-alarm tradeoff are higher for the \gls{aic} (around $-35$ dB) than for the \gls{bic} (around $-45$ dB).
These default operating values are insensitive to $\ObsAll$, $\SubAll$, $\TgtAll$ and the \gls{snr}, but would require tuning for different $\SRxAll$ as it defines the width of the hit region around the true target \gls{doa}.

%% file: Text/6-Conclusion.tex
\section{Conclusion} \label{sec:conclusion}
This paper studies the joint estimation of the number of targets and their \glspl{doa} by integrating the \gls{itc} model-order selection into the \gls{ols} greedy search.
Three methodologies are proposed: the disjoint rank-based, the joint selection-based, and the hybrid rank and selection-based \gls{itc}-\gls{ols} algorithms.
Each algorithm was derived under both the \gls{aic} and the \gls{bic} frameworks.
Numerical simulations demonstrate that the proposed hybrid \gls{itc}-\gls{ols} achieves the best overall performance by exploiting the complementarity of the rank-based and selection-based \gls{itc} components,across all considered parameters regimes.
Furthermore, \gls{aic} consistently outperforms \gls{bic} in the evaluated scenario.
A key feature of the joint and hybrid strategies is the \gls{ml} correction parameter, set to $\NoiseVarMM \approx -35$ dB for the \gls{aic} and $\NoiseVarMM \approx -45$ dB for the \gls{bic}, which controls the hit/false-alarm tradeoff at high \gls{snr} without degrading the performance at lower \gls{snr} regimes.

%% file: references.bib
@ARTICLE{5895106,
  author={Cai, T. Tony and Wang, Lie},
  journal={IEEE Transactions on Information Theory}, 
  title={Orthogonal Matching Pursuit for Sparse Signal Recovery With Noise}, 
  year={2011},
  volume={57},
  number={7},
  pages={4680-4688},
  doi={10.1109/TIT.2011.2146090}}

@article{LIANG2017165,
title = {Theoretical stopping criteria guided Greedy Algorithm for Compressive Cooperative Spectrum Sensing},
journal = {Computer Communications},
volume = {111},
pages = {165-175},
year = {2017},
issn = {0140-3664},
doi = {https://doi.org/10.1016/j.comcom.2017.08.007},
url = {https://www.sciencedirect.com/science/article/pii/S014036641730868X},
author = {Wei-Jie Liang and Tsung-Hsun Chien and Chun-Shien Lu}
}

@article{12047,
author = {Roy, Shirshendu and Acharya, Debiprasad P. and Sahoo, Ajit K.},
title = {Fast {OMP} algorithm and its {FPGA} implementation for compressed sensing-based sparse signal acquisition systems},
journal = {IET Circuits, Devices \& Systems},
volume = {15},
number = {6},
pages = {511-521},
doi = {https://doi.org/10.1049/cds2.12047},
url = {https://ietresearch.onlinelibrary.wiley.com/doi/abs/10.1049/cds2.12047},
eprint = {https://ietresearch.onlinelibrary.wiley.com/doi/pdf/10.1049/cds2.12047},
year = {2021}
}

@article{chen_orthogonal_1989,
	title = {Orthogonal least squares methods and their application to non-linear system identification},
	volume = {50},
	issn = {0020-7179},
	doi = {10.1080/00207178908953472},
	number = {5},
	urldate = {2024-12-13},
	journal = {International Journal of Control},
	author = {Chen, S. and Billings, S. A. and Luo, W.},
	month = nov,
	year = {1989},
	note = {Publisher: Taylor \& Francis
\_eprint: https://doi.org/10.1080/00207178908953472},
	pages = {1873--1896},
}

@inproceedings{blumensath_difference_2007,
	title = {On the {Difference} {Between} {Orthogonal} {Matching} {Pursuit} and {Orthogonal} {Least} {Squares}},
	urldate = {2024-12-13},
	author = {Blumensath, T. and Davies, M.},
	month = mar,
	year = {2007},
}

@article{pesavento_three_2023,
	title = {Three {More} {Decades} in {Array} {Signal} {Processing} {Research}: {An} optimization and structure exploitation perspective},
	volume = {40},
	issn = {1558-0792},
	shorttitle = {Three {More} {Decades} in {Array} {Signal} {Processing} {Research}},
	doi = {10.1109/MSP.2023.3255558},
	number = {4},
	urldate = {2024-12-13},
	journal = {IEEE Signal Processing Magazine},
	author = {Pesavento, Marius and Trinh-Hoang, Minh and Viberg, Mats},
	month = jun,
	year = {2023},
	pages = {92--106},
}

@article{CHEN2019331,
title = {A blind stopping condition for orthogonal matching pursuit with applications to compressive sensing radar},
journal = {Signal Processing},
volume = {165},
pages = {331-342},
year = {2019},
issn = {0165-1684},
doi = {https://doi.org/10.1016/j.sigpro.2019.07.022},
url = {https://www.sciencedirect.com/science/article/pii/S0165168419302750},
author = {Shengyao Chen and Zhiyong Cheng and Chao Liu and Feng Xi}
}

@ARTICLE{10979421,
  author={Willame, Martin and Monnoyer, Gilles and Yildirim, Hasan Can and Horlin, François and Louveaux, Jérôme},
  journal={IEEE Signal Processing Letters}, 
  title={Multi Target Localization With Block Orthogonal Least Squares for Multistatic {MIMO} Radars}, 
  year={2025},
  volume={32},
  number={},
  pages={1990-1994},
  doi={10.1109/LSP.2025.3565168}}

@ARTICLE{7543,
  author={Ziskind, I. and Wax, M.},
  journal={IEEE Transactions on Acoustics, Speech, and Signal Processing}, 
  title={Maximum likelihood localization of multiple sources by alternating projection}, 
  year={1988},
  volume={36},
  number={10},
  pages={1553-1560},
  doi={10.1109/29.7543}}

@ARTICLE{1164557,
  author={Wax, M. and Kailath, T.},
  journal={IEEE Transactions on Acoustics, Speech, and Signal Processing}, 
  title={Detection of signals by information theoretic criteria}, 
  year={1985},
  volume={33},
  number={2},
  pages={387-392},
  doi={10.1109/TASSP.1985.1164557}}

@ARTICLE{1311138,
  author={Stoica, P. and Selen, Y.},
  journal={IEEE Signal Processing Magazine}, 
  title={Model-order selection: a review of information criterion rules}, 
  year={2004},
  volume={21},
  number={4},
  pages={36-47},
  doi={10.1109/MSP.2004.1311138}}

@ARTICLE{8498082,
  author={Ding, Jie and Tarokh, Vahid and Yang, Yuhong},
  journal={IEEE Signal Processing Magazine}, 
  title={Model Selection Techniques: An Overview}, 
  year={2018},
  volume={35},
  number={6},
  pages={16-34},
  doi={10.1109/MSP.2018.2867638}}

@inproceedings{Akaike1973InformationTA,
author = {Akaike, H.},
title = {Information theory and an extension of the maximum likelihood principle},
booktitle = {In B. N. Petrov \& F. Csáki (Eds.), 2nd international symposium on information theory (pp. 267–281). Budapest, Hungary: Akadémia Kiadó},
year = {1973}
}

@article{Cavanaugh2019AkaikeIT,
author = {Cavanaugh, Joseph E. and Neath, Andrew A.},
title = {The Akaike information criterion: Background, derivation, properties, application, interpretation, and refinements},
year = {2019},
issue_date = {May/June 2019},
publisher = {John Wiley \& Sons, Inc.},
address = {USA},
volume = {11},
number = {3},
issn = {1939-5108},
url = {https://doi.org/10.1002/wics.1460},
doi = {10.1002/wics.1460},
journal = {WIREs Comput. Stat.},
month = apr,
numpages = {11}
}

@article{Schwarz1978EstimatingTD,
 ISSN = {00905364, 21688966},
 URL = {http://www.jstor.org/stable/2958889},
 author = {Gideon Schwarz},
 journal = {The Annals of Statistics},
 number = {2},
 pages = {461--464},
 publisher = {Institute of Mathematical Statistics},
 title = {Estimating the Dimension of a Model},
 urldate = {2026-01-08},
 volume = {6},
 year = {1978}
}

@article{Neath2012BayesianIC,
author = {Neath, Andrew A. and Cavanaugh, Joseph E.},
title = {The Bayesian information criterion: background, derivation, and applications},
year = {2012},
issue_date = {March/April 2012},
publisher = {John Wiley \& Sons, Inc.},
address = {USA},
volume = {4},
number = {2},
issn = {1939-5108},
url = {https://doi.org/10.1002/wics.199},
doi = {10.1002/wics.199},
journal = {WIREs Comput. Stat.},
month = feb,
pages = {199–203},
numpages = {5}
}

@inproceedings{Richards2013PrinciplesOM,
  title={Principles of Modern Radar: Basic Principles},
  author={Mark A. Richards and James A. Scheer and William A. Holm},
  year={2013},
  url={https://api.semanticscholar.org/CorpusID:114114032}
}

@article{Kay1993FundamentalsOS,
  title={Fundamentals of statistical signal processing: estimation theory},
  author={Steven M. Kay},
  journal={Technometrics},
  year={1993},
  volume={37},
  pages={465},
  url={https://api.semanticscholar.org/CorpusID:120930025}
}

@ARTICLE{9477585,
  author={Storrer, Laurent and Yildirim, Hasan Can and Crauwels, Morgane and Copa, Evert I. Pocoma and Pollin, Sofie and Louveaux, Jérôme and De Doncker, Philippe and Horlin, François},
  journal={IEEE Sensors Journal}, 
  title={Indoor Tracking of Multiple Individuals With an 802.11ax {Wi-Fi}-Based Multi-Antenna Passive Radar}, 
  year={2021},
  volume={21},
  number={18},
  pages={20462-20474},
  doi={10.1109/JSEN.2021.3095675}}

@Inbook{Kuhn2010,
author="Kuhn, Harold W.",
editor="J{\"u}nger, Michael
and Liebling, Thomas M.
and Naddef, Denis
and Nemhauser, George L.
and Pulleyblank, William R.
and Reinelt, Gerhard
and Rinaldi, Giovanni
and Wolsey, Laurence A.",
title="The Hungarian Method for the Assignment Problem",
bookTitle="50 Years of Integer Programming 1958-2008: From the Early Years to the State-of-the-Art",
year="2010",
publisher="Springer Berlin Heidelberg",
address="Berlin, Heidelberg",
pages="29--47",
isbn="978-3-540-68279-0",
}

@misc{salama2025DoA,
      title={Direction of Arrival Estimation: A Tutorial Survey of Classical and Modern Methods}, 
      author={Amgad A. Salama},
      year={2025},
      eprint={2508.11675},
      archivePrefix={arXiv},
      primaryClass={eess.SP},
      url={https://arxiv.org/abs/2508.11675}, 
}

@book{Foutz2008DoA,
author = {Jeffrey Foutz, Andreas Spanias, Mahesh K. Banavar},
title = {Narrowband Direction of Arrival Estimation for Antenna Arrays},
publisher = {Springer Cham},
year = {2008},
doi = {10.1007/978-3-031-01537-3},
address = {},
edition   = {},
eprint = {https://link.springer.com/book/10.1007/978-3-031-01537-3}
}

@ARTICLE{11090080,
  author={},
  journal={IEEE Std 802.11be-2024 (Amendment to IEEE Std 802.11-2024, as amended by 802.11bh-2024)}, 
  title={{IEEE} Standard for Information technology--Telecommunications and information exchange between systems Local and metropolitan area networks--Specific requirements - Part 11: Wireless {LAN} Medium Access Control ({MAC}) and Physical Layer ({PHY}) Specifications Amendment 2: Enhancements for Extremely High Throughput ({EHT})}, 
  year={2025},
  volume={},
  number={},
  pages={1-1020},
  doi={10.1109/IEEESTD.2024.11090080}}

@article{Youden1950Index,
author = {Youden, W. J.},
title = {Index for rating diagnostic tests},
journal = {Cancer},
volume = {3},
number = {1},
pages = {32-35},
doi = {https://doi.org/10.1002/1097-0142(1950)3:1<32::AID-CNCR2820030106>3.0.CO;2-3},
year = {1950}
}

@article{MartinezCamblor2019,
  author = {Mart{\'\i}nez-Camblor, Pablo and Pardo-Fern{\'a}ndez, Juan Carlos},
  title = {The Youden Index in the Generalized Receiver Operating Characteristic Curve Context},
  journal = {International Journal of Biostatistics},
  year = {2019},
  volume = {15},
  issue = {1},
  doi = {10.1515/ijb-2018-0060}
}

@ARTICLE{1143830,
  author={Schmidt, R.},
  journal={{IEEE} Transactions on Antennas and Propagation}, 
  title={Multiple emitter location and signal parameter estimation}, 
  year={1986},
  volume={34},
  number={3},
  pages={276-280},
  doi={10.1109/TAP.1986.1143830}}
